# Nanoscale Control over Optical Dislocations

**Evgeny Ostrovsky, Kobi Cohen, Bergin Gjonaj and Guy Bartal**

*Technion – Israel Institute of Technology, Haifa 32000, Israel*
*Author e-mail address: ostrovsky@campus.technion.ac.il*

**Abstract:** High-order optical vortices are inherently unstable, as they tend to split up under perturbation to a series of vortices with unity charge. Control over the perturbation opens up a new degree of freedom to control and tune their location in a 2D-space, with immediate implications on trapping, light-matter interactions and super-resolution imaging. Here, we present a continuous nano-scale tuning of the spatial location of plasmonic vortices on metal-air interface achieved by varying the polarization state of the light coupled to the surface plasmons through a spiral slit. We demonstrate such a control at nano-scale resolution using phase-resolved near-field microscopy.

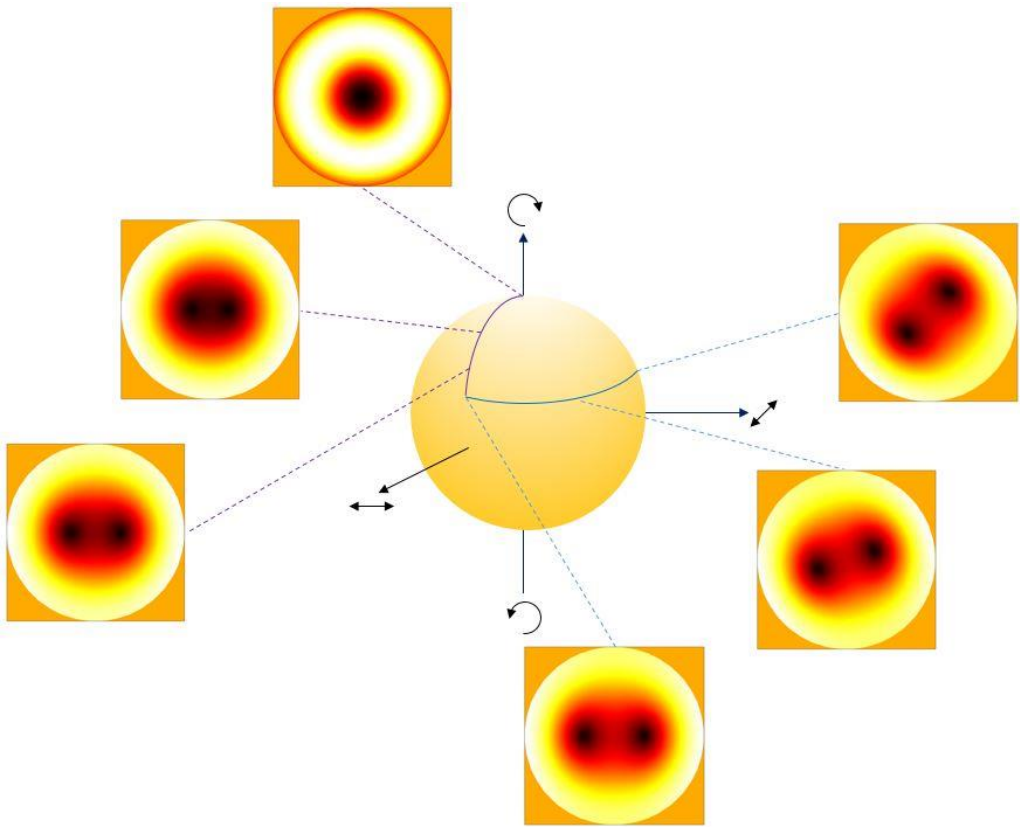

## Introduction

Optical Dislocations [1] are singularity points in Electromagnetic fields having zero intensity in the center, accompanied by azimuthal phase variations. An important class of beams with such wave-fronts are Optical Vortices (OV), which are known to carry an Orbital Angular Momentum (OAM) of $l\hbar$ per photon where $l$ is an integer defined as the topological charge of the vortex. These OVs OAM can be converted into a mechanical torque [2] so as to be used to trap and manipulate nanometer-sized particles [3, 25, 26], can transfer information via optical communication [4] and enable high capacity wireless communications [5], spatial resolution beyond diffraction limit [6, 7] and small scale optical communication between chips [8]. In addition, it is expected to assist in disentanglement of spatial modes of the electromagnetic field and thus enhance the quantum information capacity [9, 10].

Some of the above-mentioned applications and, in particular, those concerned with light-matter interaction, require significant reduction of the vortex size beyond the diffraction limit of free space. Such a reduction in the two spatial dimensions is accompanied by confinement in the third one, defining a subclass of 2D OVs. These can enhance the applicability of OVs especially in the field of light matter interaction, providing new schemes involving OAM and redefining selection rules for quantum transitions [11, 12, 13] and phenomenons such as Fano Resonances [20]. Achieving optimal light-matter interaction of that kind along with deeper investigation requires a precise control of the OVs properties, such as their accurate location in the 2D space. This is typically evaded in structurally generated OVs or polarization singularities that are intrinsically controlled by nanofabrication and scattering [27] or plasmonic vortices generated by specifically-designed coupling slits [14, 15, 16, 17].

Interestingly, such degree of control can be achieved by utilizing the inherent instability of high order OVs, which are sensitive to imperfections and tend to break under perturbations - into a series of lower order OVs [19]. While this property is often considered a drawback of OVs which has witnessed many attempts to correct the imperfection (see, e.g. [24]), a controlled coherent perturbation may lead to precise control of the new dislocations created, thereby controlling the location of the (lower order) OVs.

Here, we demonstrate a nanoscale-precision tuning of optical dislocations by controlling the splitting of high-order plasmonic Bessel beams. This is demonstrated by varying the polarization state of a plane wave coupled to Surface Plasmons Polaritons (SPPs) through a helical slit which in turn, determines the relative amplitude and phase between two Bessel beams of different order. These two degrees of freedom, provided by the two degrees of freedom of the polarization state (best expressed by the Poincare Sphere) results in controlled interference between the two different-order Bessel modes which determines the location of the generated dislocations at nanoscale precision.

## Mathematical background

OVs can be created either at the focal plane of helical beams, or by propagating surface or guided modes on a two-dimensional plane [14, 17, 21, 22]. The first type is usually described as Laguerre-Gaussian (LG) beams [2], which is a self-similar solution of the paraxial wave equation, while the surface OVs are two-dimensional, decaying exponentially in the direction normal to the interface. Such 2D OVs were demonstrated in plasmonic structures [14, 17, 21, 22] and Silicon-plasmonic waveguides [7, 15].

Being a solution of two-dimensional space, surface OVs take the form of cylindrical harmonics with the radial part given by Bessel functions of the first kind:

$$E(r,\theta) = J_m(kr)e^{im\theta} \qquad (1)$$

Where $J_m$ is the $m$th-order Bessel function of the first kind, $r$ and $\theta$ are the cylindrical coordinate set, $k$ is the wave number in the surface and the order $m$ is the number of periods the phase completes in the azimuthal direction within a full circle, aka the topological charge of the vortex.

The surface modes can be excited by means of momentum matching between free propagating light and the guided mode, for example, plasmons on a metal-dielectric interface [28] or by grating or slit coupling. When the slit width is much smaller than the wavelength of the incident light, it practically acts as a polarizer, which couples the electric field components normal to the slit/grating into the surface mode. This leads to an interplay between the polarization states of the incident light to the phase variations along the slit ("source" for the guided modes) which depends on both the shape of the slit and the illumination's polarization, providing an important degree of freedom in shaping high order of Surface Bessel beams [14].

Consider a Helical slit carved in metal to couple incident radiation into SPP modes (Fig.6). This slit is identified by the separation distance between successive turnings which, for plasmons, is $d = l\lambda_{spp}$, $\lambda_{spp}$ is the wavelength of the SPP and $l$ is an integer which defines the geometrical Topological Charge of the lens. Coupling circularly polarized light, i.e. Angular Momentum of $\sigma = \pm 1$, into the guided modes via such a lens results in a Bessel beam with a topological charge of the sum of the light's OAM and the lens' geometrical topological charge i.e., $l \pm 1$ respectively:

$$E(r,\theta) = J_{(l\pm1)}(kr)e^{i(l\pm1)\theta} \qquad (2)$$

Where $l$ is the lens' geometrical topological charge.

This property has been demonstrated in numerous works, starting from $l = 1$ slit resulting in Bessel beams of orders 0 and 2 [14] up to higher order OAMs [16, 17, 15]. These observations have indeed shown a ring whose radius increases with the Bessel order, but at the same time, did not provide evidence to high topological charge singularity; it either lacked the phase information, or showed that close to the center the high-order OV tend to break into a series of single-unity OVs [19, 17]. This breakdown is attributed to the lack of pure state, namely, the high order Bessel beams are "contaminated" by noise, background or leftovers from the lower order Bessel beam. In plasmonic Bessel beams, this leftover stems from fabrication imperfections but also from non-pure polarization state. As this phenomenon occurs close to the singularity where the intensity is low, it is typically overlooked by the intensity pattern and can only be observed in the phase structure (see simulation in [14] and phase-resolved measurement in [17]).

Surprisingly, this seemingly drawback can be utilized to control the splitting of the higher order Bessel beam, i.e., the location of the newly-generated single-unity OVs by merely controlling the polarization state of the excitation beam, achievable e.g., using half- and quarter- waveplates.

In fact, as any polarization state can be expressed as super-position of the two circular polarizations, the naturally-polarizing slit couplers directly transform the polarization state to super-position of two Bessel beams at orders l+1 and l-1 where the amplitude and the phase relations between the two circular polarizations is preserved in the ratio between the Bessel states. Here, we use this property to demonstrate the precision control over these plasmonic dislocations in the following way:

General polarization state can be expressed as a superposition of the two circular polarizations:

$$\sigma_{+1} + |a_0|e^{i\varphi_0}\sigma_{-1} \tag{3}$$

Where $a_0$ and $\varphi_0$ represents the relative amplitude ratio and the phase between the two circular polarizations respectively.

The naturally-polarizing slit couplers directly transform the polarization state to super-position of two Bessel beams of orders $l+1$ and $l-1$:

$$E(r,\theta) = J_{l+1}(kr)e^{i(l+1)\theta} + |a_0|e^{i\varphi_0}J_{l-1}(kr)e^{i(l-1)\theta} \tag{4}$$

Where $a_0$ and $\varphi_0$ (the same coefficients of the polarization state) now represent the relative amplitude ratio and phase difference of the two Bessel beams respectively. These coefficients can be controlled by a Quarter ($\lambda/4$) and Half ($\lambda/2$) wave plates:

$$|a_0| = f_{a_0}(\vartheta_{\lambda/4}, \vartheta_{\lambda/2}); \varphi_0 = f_{\varphi_0}(\vartheta_{\lambda/4}, \vartheta_{\lambda/2}) \tag{5}$$

Where $\vartheta_{\lambda/4}, \vartheta_{\lambda/2}$ are the orientation angles of the Quarter wave-plate and Half wave plates respectively, relative to the linear polarization of the incident beam and $f_{a_0}, f_{\varphi_0}$ are functions which can be derived simply by analyzing the polarization state after the linear polarized light passing the two plates:

$$f_{a_0}(\vartheta_{\lambda/4}, \vartheta_{\lambda/2}) = \left|\frac{f+ie}{f-ie}\right|; f_{\varphi_0}(\vartheta_{\lambda/4}, \vartheta_{\lambda/2}) = \arg\left(\frac{f+ie}{f-ie}\right)$$

$$f = \frac{d+ic}{2i}; e = \frac{d-ic}{-2i}$$

$$d = a\cos\vartheta_{\lambda/2} + b\sin\vartheta_{\lambda/2}; c = a\sin\vartheta_{\lambda/2} - b\cos\vartheta_{\lambda/2}$$

$$a = \sin^2\vartheta_{\lambda/4} + i\cos^2\vartheta_{\lambda/4}; b = \frac{1}{2}\sin2\vartheta_{\lambda/4}(1-i)$$

While the amplitude ratio controls the distances between the OVs, the phase difference controls the orientation axis of the OVs.

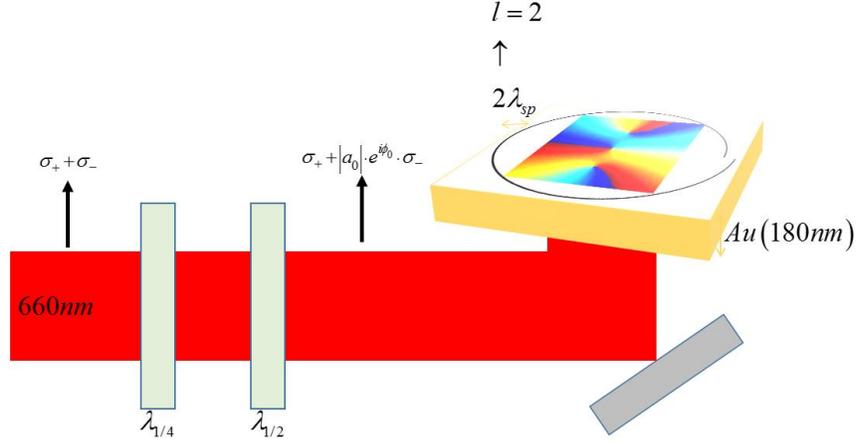

**Fig. 1 Polarization state control.** A (linearly polarized) laser beam propagates through a $\lambda/2$ and a $\lambda/4$ plates which provide complete control over the polarization state. The beam is then incident upon a metal-dielectric interface coupled by the spiral slit ("plasmonic lens"). The polarization state is translated to two Bessel beams of different orders that create the multiple vortices with controlled position. The full field distribution is mapped by phase-resolved s-NSOM showing both the amplitude and phase at 15nm resolution.

## Experiment

The light, carrying a polarization state determined by the two waveplates [figure 1], is then launched through an Archimedean spiral slit with topological charge of $l = 1, 2$ (two different experiments) to couple the incoming light to Plasmonic mode thus provides the control over the OVs location. We use a scattering type near-field scanning optical microscope (s-NSOM) that enables phase-resolved mapping of the near field on the metal surface via pseudo-heterodyne interferometric detection [23].

We demonstrate the control over the distance between the OVs (the radial position) by changing the relative amplitude between the circular polarizations, i.e. mainly changing the orientation of the $\lambda/4$ plate. Control over the rotation axis of the OVs (azimuthal position) is achieved by changing the relative phase i.e. mainly changing the orientation of the $\lambda/2$ plate.

The radial position control is demonstrated in Fig. 2 and Fig. 3 where the topological charge of the coupling slit generates plasmonic Bessel beams of orders 0 and 2 for $L = 1$ (Fig. 2), or orders 1 and 3 for $L = 2$ (Fig. 3). Using the $\lambda/4$ we control the amplitude ratio between these two Bessel beams hence the distance of the two newly-generated OVs from the center. Both amplitude and phase are shown, where the phase singularities are marked by black circles and correspond to the dark spots in the amplitude.

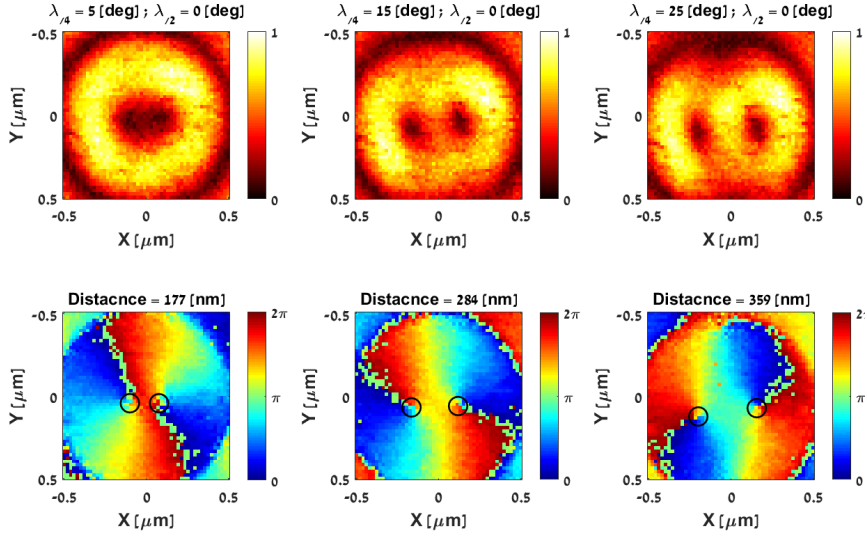

Fig. 2 Near-field mapping of Nanoscale Control over Optical Dislocations for $L = 1$. Control of distance between OVs; The orientations of the $\lambda/4$ and $\lambda/2$ plates are indicated above the amplitude graphs; The upper row is the amplitude and the lower row is the phase of the near field; the distance between the OV is indicated above the phase graphs; OVs are marked in the phase graphs.

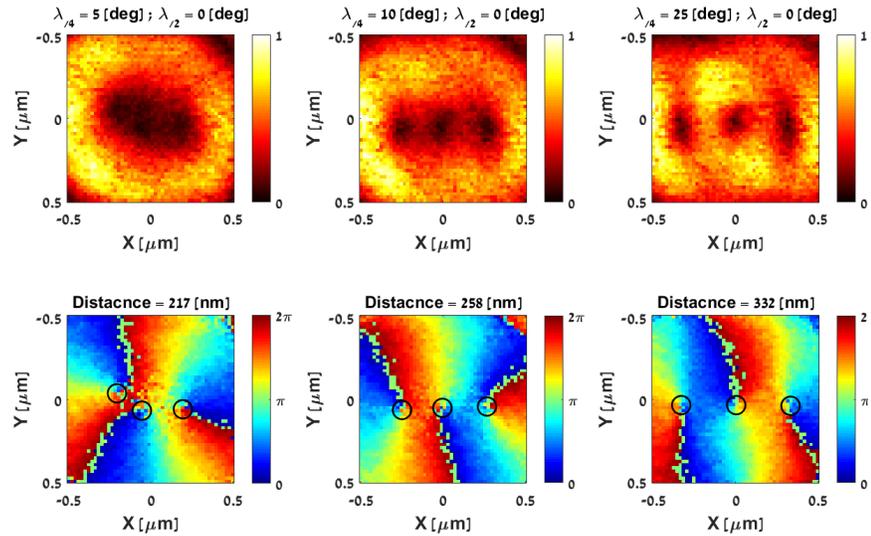

Fig. 3 Near-field mapping of Nanoscale Control over Optical Dislocations for $L = 2$. Control of distance between OVs; The orientations of the $\lambda/4$ and $\lambda/2$ plates are indicated above the amplitude graphs; The upper row is the amplitude and the lower row is the phase of the near field; the distance between OV is indicated above the phase graphs; OVs are marked in the phase graphs.

Adding the $\lambda/2$ plate provides additional control over the relative phase between the two, resulting in a rotational degree of freedom for the two OVs (Fig. 4 and Fig. 5). Namely, the two degrees of freedom in the polarization state, controlled by the waveplates are translated to two-dimensional position control for the optical (plasmonic) dislocations with nanoscale precision.

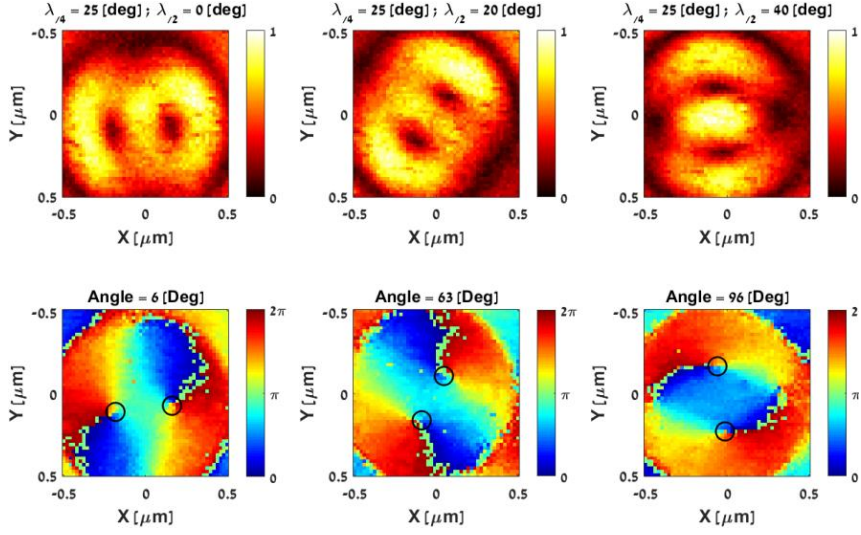

**Fig. 4 Near-field mapping of Nanoscale Control over Optical Dislocations for $L = 1$.** Control of rotation axis of OVs; The orientations of the $\lambda/4$ and $\lambda/2$ plates are indicated above the amplitude graphs; The upper row is the amplitude and the lower row is the phase of the near field; the angle of the rotation axis of the OVs is indicated above the phase graphs; OVs are marked in the phase graphs.

**Fig. 5 Near-field mapping of Nanoscale Control over Optical Dislocations for $L = 2$.** Control of rotation axis of OVs; The orientations of the $\lambda/4$ and $\lambda/2$ plates are indicated above the amplitude graphs; The upper row is the amplitude and the lower row is the phase of the near field; the angle of the rotation axis of the OVs is indicated above the phase graphs; OVs are marked in the phase graphs.

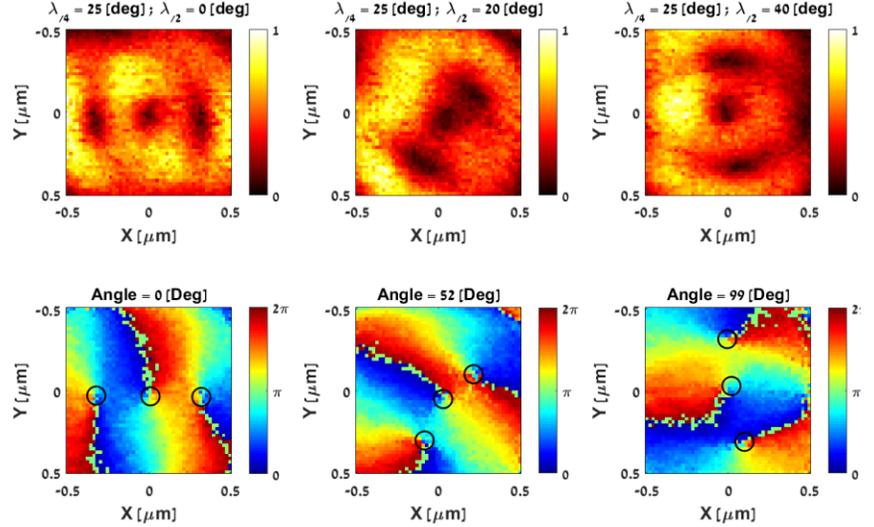

Note that slit with $L = 1$ generates lower order of 0, resulting in no OV in the center. In contrast, with slit of $L = 2$ the lower order is 1 and therefore an additional 1st order OV is present in the center. Note that when the OV get close (weak low-order Bessel as a result of polarization close to circular) the observed pattern becomes somewhat asymmetric. This is a result of a slight mismatch between the coupling spiral slit to the plasmonic wavelength due to imperfect fabrication, also hindering excitation and generation of pure high-order OVs.

## Conclusion

In conclusion, we demonstrated complete control over the location of OVs in 2D space at nanoscale resolution by simply changing the polarization state of the incident light. Such control can be extended to other 2D system, e.g. short-wavelength hybrid Si-plasmonic waveguides [17], which downscale the size of the vortices so as to allow precise control over the location of nano-OV that could be utilized to demonstrate Orbital Angular Momentum based light-matter interaction.

## Methods

**Lens fabrication**

The sample is a deposition of thin gold film of 180 nm thick atop of a glass substrate of 200 $[\mu m]$ thick. The plasmonic lens was fabricated using a Dual-beam focused ion beam (FEI Strata 400s) by milling the thin slits from the metal side. As a Plasmonic lens with topological charge of $l = 2$ we use a slit with a shape of an Archimedean spiral with a gap of $l = 2\lambda_{sp}$ where $\lambda_{sp} = 633\ [nm]$ is the wavelength of a SPPs in the Gold-Air interface and the average diameter of the lens is 16 $[\mu m]$.

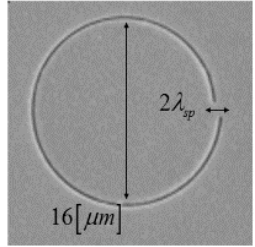

**Fig. 6 Plasmonic lens.** A Scanning Electron Microscope (SEM) image of a Plasmonic lens.

**Near field measurements**

We used a scattering (apertureless) Scanning Near-field Optical Microscope (s-SNOM, Neaspec ltd.) for the measurement of the near field signal, both amplitude and phase. The measurement setup is illustrated in Fig. 7. A 660 nm continuous wave (CW) semiconductor laser beam was weakly focused to a 50μm spot and thus illuminates the sample from the glass side (from below) aka transmission mode. The sample placed on a moving stage, whereas a metallic (Pt\Ir) AFM tip scatters predominantly the out-of-plane electric field component from the Silicon side (from above) into a detector. The tip vibrates at amplitude of approximately 27 nm and frequency of about $\Omega = 250$ MHz. The incoming laser beam was split into two optical paths for an interferometric pseudo-heterodyne detection; a beam modulated by a vibrating mirror with a frequency of $M = 600$ Hz was interfered with the signal scattered from the sample to reconstruct the full E-field information including the phase. Therefore, the detected signal is demodulated at higher harmonics $n\Omega + M$ in order to suppress the background signal scattered from the tip [23]. The resolution ability of the s-NSOM is determined by the size of the tip apex, which is, in our case, between 10-20 nm.

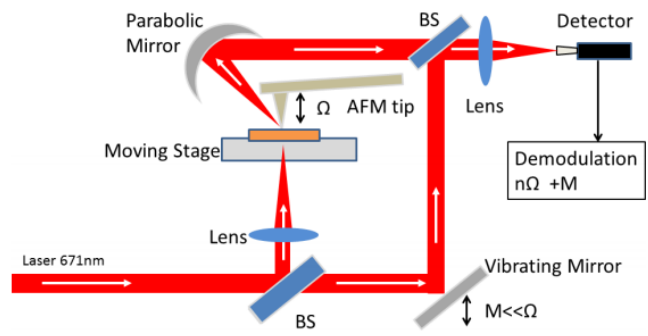

**Fig. 7 NSOM measurement system.** S-NSOM system used for a full phase-resolved field (both the amplitude and phase) distribution mapping.